\documentclass[cits]{PoS}
\usepackage{amssymb,amsmath}

\newcommand{\lsim}{
\mathrel{\hbox{\rlap{\hbox{\lower4pt\hbox{$\sim$}}}\hbox{$<$}}}}

\newcommand{\gsim}{
\mathrel{\hbox{\rlap{\hbox{\lower4pt\hbox{$\sim$}}}\hbox{$>$}}}}

\title{\boldmath Nuclear modification factor in $p+Pb$ collisions at LHC and saturation }

\ShortTitle{Nuclear modification factor in $p+pb$ collisions at LHC and saturation }

\author{\speaker{Amir H. Rezaeian}
\\
Institut f\"ur Theoretische Physik, Universit\"at Regensburg, 93040 Regensburg, Germany\\
Departamento de F\'\i sica y Centro de Estudios Subat\'omicos,\\ Universidad T\'ecnica Federico Santa Mar\'\i a, Casilla 110-V, Valpara\'\i so, Chile\\
        E-mail: \email{amir.rezaeian@usm.cl}}

\abstract{ We provide
 predictions for the nuclear modification factor $R_{pA}$ for pions
 and direct photon production in $p+A$ collisions at LHC energy at
 midrapidity within different saturation models fitted to HERA
 data. In our approach we have no free parameters to adjust and all
 model parameters are fitted to other reactions. Our approach gives a
 rather good description of PHENIX data for $R_{pA}$
 for pions. We show that, in various saturation models, the pion Cronin enhancement is replaced by
 a moderate suppression at LHC energy at midrapidity due to gluon
 shadowing effects. However, Cronin
 enhancement of direct photons can survive at LHC energy
 in models with a larger saturation scale. We show that both
 shadowing and saturation effects are important at LHC in $p+A$
 collisions and give rise to a rather sizable effect in the nuclear
 modification factor $R_{pA}$. Therefore, a precise measurement
 of $p+A$ collisions at LHC is crucial in order to understand the
 underlying dynamics of heavy ion collisions.  }

\FullConference{The 2009 Europhysics Conference on High Energy Physics,\\
		 July 16 - 22 2009\\
		 Krakow, Poland}

\begin{document}


It is believed that data for $p+p$ and $p+A$ collisions
can be indirectly translated into initial state effects in nuclear collisions. Therefore, in order to interpret jet-quenching, it seems mandatory to have a precise
and firm understanding of the Cronin, shadowing and saturation effects in $p+A$ collisions \cite{palhc}.
The nuclear modification (Cronin) factor $R_{pA}$ is defined as ratio of $p+A$ to
$p+p$ cross-sections normalized to the average number of binary nucleon
collisions. In our approach, the Cronin effect originates from initial-state
broadening of the transverse momentum of a projectile parton
interacting coherently with a nuclear medium \cite{RS}. The invariant
cross-section of hadron and direct photon production in $p+A$ (and $p+p$)
collisions can be calculated via the light-cone color-dipole
factorization scheme \cite{RS}, see also Refs.~\cite{me1}. One should note that the multiple parton
interactions that lead to gluon shadowing are also the source of gluon
saturation. The saturation effects have been already included in
the color-dipole cross-section fitted to DIS data. In order to avoid
double counting, we incorporate the nuclear shadowing effect within the
same color-dipole formulation. Details of calculations can be found in Ref.~\cite{RS}.
\begin{figure}[t]
\center{\begin{minipage}{6cm}
\includegraphics[width=\textwidth]{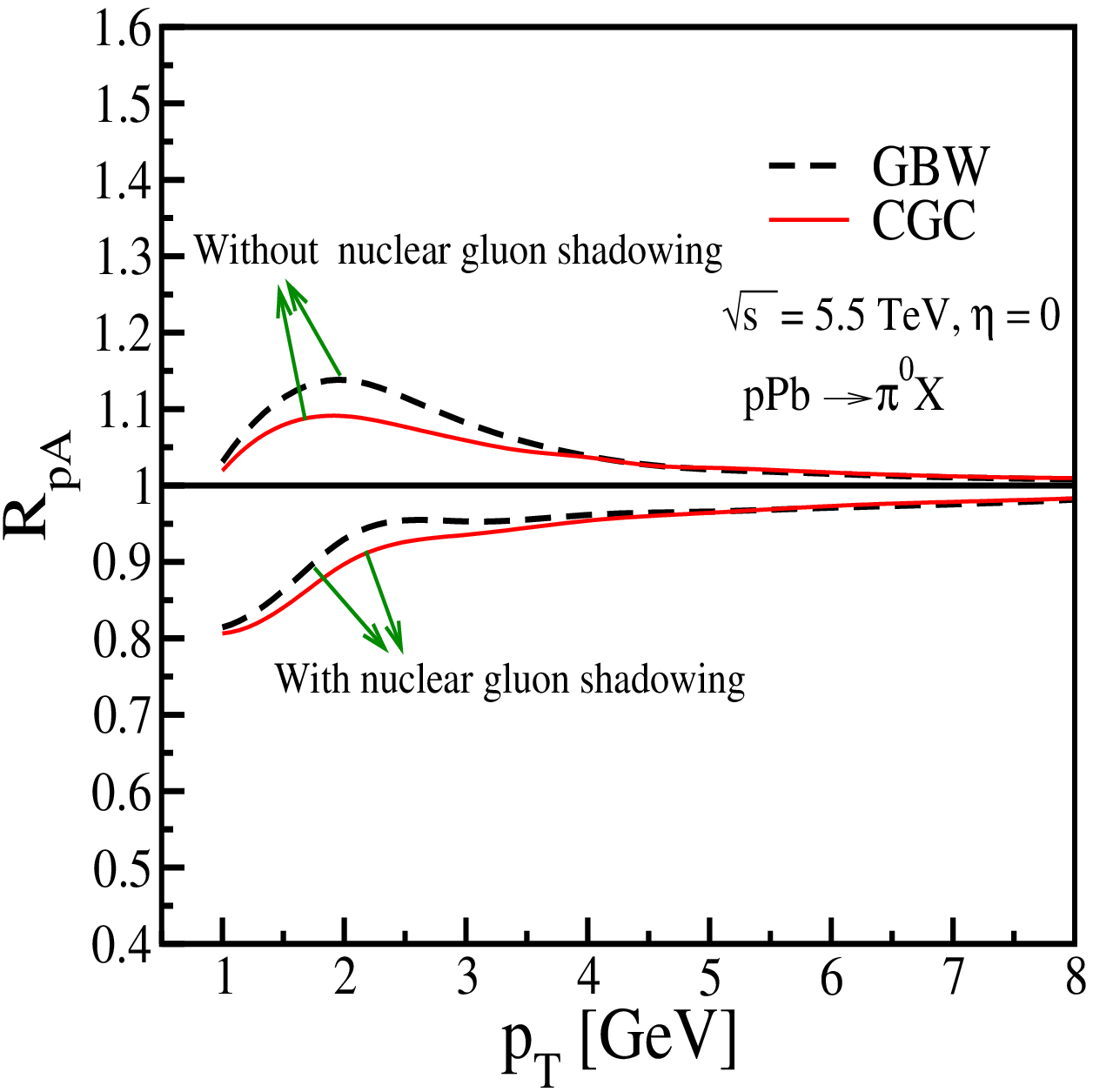}
\end{minipage}
\hspace{1cm}
\begin{minipage}{6cm}
\includegraphics[width=\textwidth]{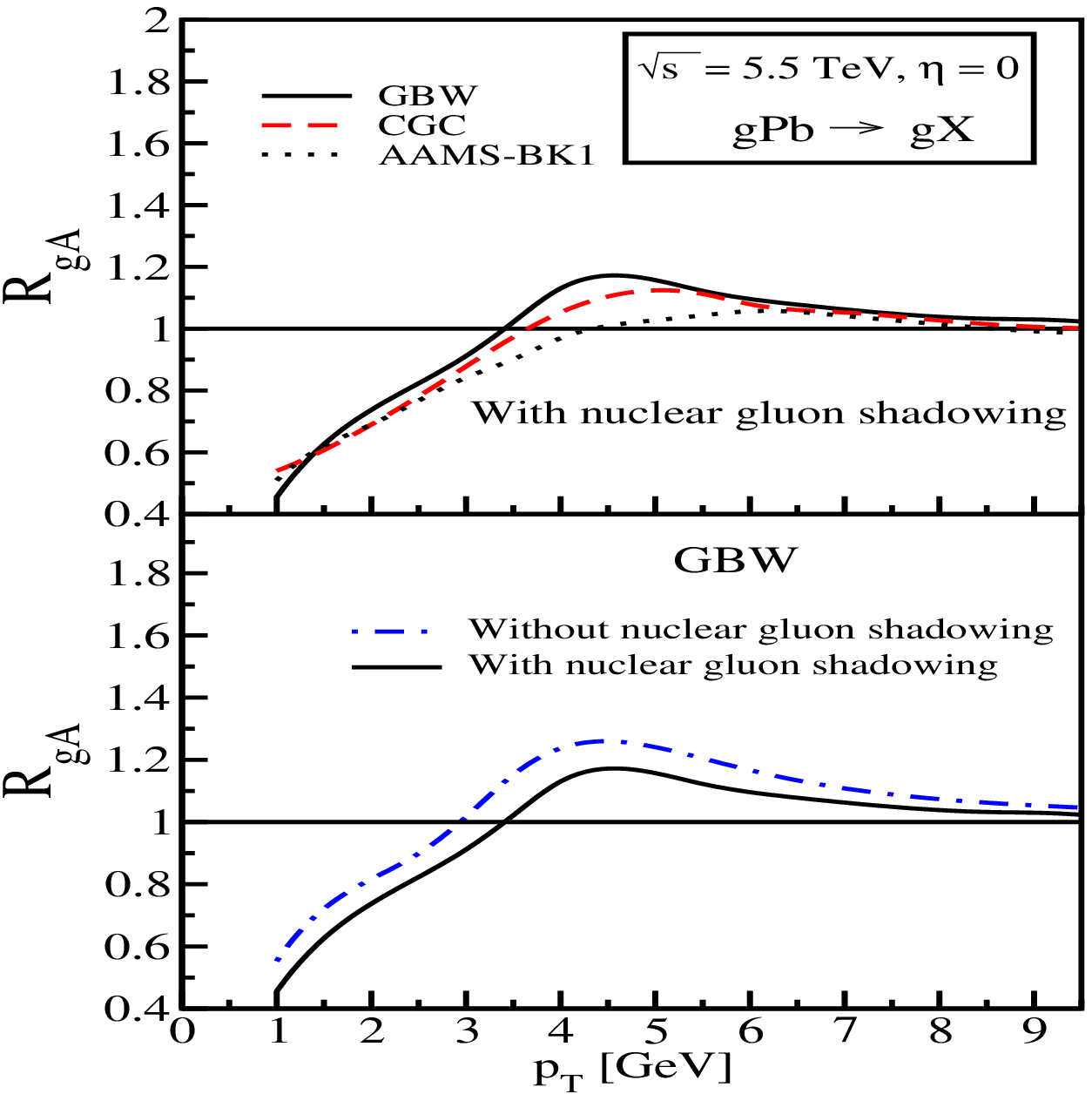}
\end{minipage}}
\hfill
\caption{Left:Nuclear modification factor $R_{pA}$ for pion production at LHC at midrapidity in minimum bias proton-lead collisions 
within the CGC and GBW color dipole models. Right: $R_{gA}$ for gluon
production. In all curves in the upper panel gluon shadowing is
incorporated. In the lower panel: gluon shadowing effects at LHC for the GBW
model.}
\label{clhc1}
\end{figure}

\begin{figure}[t]
\center{\begin{minipage}{6cm}
\includegraphics[width=\textwidth]{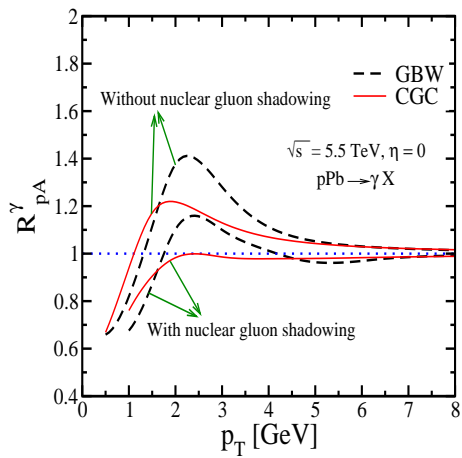}
\end{minipage}}
\caption{Same as Fig.~1(left) for direct photon production. For comparison, we also show the results with and without inclusion of gluon shadowing effects.}
\label{clhc3}
\end{figure}

In Fig.~\ref{clhc1}, we show our prediction for the nuclear
modification factor $R_{pA}$ for $\pi^0$ production at LHC at
midrapidity in minimum bias $p+A$ collisions within two very different
saturation models, namely GBW \cite{gbw} and CGC model \cite{cgc}. The CGC model is based on
the non-linear small-x Balitsky-Kovchegov equation \cite{bk}.  The saturation
scale in the phenomenological GBW model is bigger than in the CGC model (see Fig. 2 in Ref.~\cite{RS}). However, both models are able to describe deep inelastic
lepton-hadron scattering data. In Fig.~\ref{clhc1} we also show the
effect of nuclear gluon shadowing.  It is seen that the Cronin enhancement
will be replaced with moderate suppression due to nuclear gluon shadowing. It
is obvious that a bigger saturation scale leads to a larger Cronin
enhancement and works against the nuclear shadowing suppression. Note that a
larger saturation scale leads to a stronger broadening of transverse
momentum of the projectile partons and consequently it works against
nuclear shadowing. This is more obvious in Fig.~\ref{clhc1} (right, upper
panel) where we plotted the Cronin ratio for gluon production at the
LHC energy within various saturation color dipole models.  In
Fig.~\ref{clhc1} (right, lower panel) we show the effect of nuclear shadowing
within the GBW color dipole model. It is seen that both nuclear shadowing and
saturation effects are important at LHC in $p+A$ collisions and give
rise to a rather sizable effect in the nuclear modification factor
$R_{pA}$. Kharzeev {\em et al.} \cite{cro-cgc1} have shown a marked
suppression for pions at mid-rapidity at LHC in $p+A$ collisions based
on the CGC scenario. This suppression is stronger than our
prediction. Other predictions based on different approaches can be
found in Ref.~\cite{palhc}. In Fig.~\ref{clhc3}, We show our
prediction for the nuclear modification factor $R_{pA}^{\gamma}$ for
direct photon production at LHC at midrapidity in minimum bias $p+A$
collisions for two models with different saturation scale. In
comparison to pion production, the Cronin enhancement for direct
photon production seems stronger and survives within the GBW
color-dipole model which has a bigger saturation scale, even after the
inclusion of nuclear shadowing suppression effects. Similar to pion
production, the Cronin enhancement for direct photon production is
bigger in a model with a larger saturation scale.  Within the CGC
model both pion and direct photon enhancement at RHIC will be replaced
by suppression at LHC.


\section*{Acknowledgments}
Warmest thanks are given to my collaborators Boris Kopeliovich and Andreas Sch\"afer for their contributions to the work summarised here. 
A. R. acknowledges the financial support from the Alexander von Humboldt foundation, BMBF (Germany), Conicyt Programa
Bicentenario PSD-91-2006 and Fondecyt grants 1090312 (Chile).

\end{document}